\documentclass[preprint,prd,showpacs,preprintnumbers,amsmath,amssymb,floatfix]{revtex4}
\usepackage{amsmath}
\usepackage{lipsum}
\usepackage{graphicx}
\usepackage{dcolumn}
\usepackage{bm}
\usepackage{ulem} 
\usepackage[usenames]{color}
\usepackage{epstopdf}
\usepackage{epsfig}
\usepackage{float}
\usepackage{subfigure}
\usepackage{multirow}
\usepackage[version=3]{mhchem}

\usepackage{soul}

\newcommand{\nc}{\newcommand}       
\nc{\vc}[1] {\mbox{\boldmath $#1$}} 
\nc{\del}       {\partial}              
\nc{\bra}       {\langle}               
\nc{\ket}       {\rangle}               
\nc{\bras}[1]   {\langle #1|}           
\nc{\kets}[1]   {|#1\rangle}            
\nc{\mapleft}[1]{           
 \smash{\mathop{\,          %
  \hbox to 1.5cm{\rightarrowfill}\, }\limits_{#1}}}
\nc{\beq}     {\begin{eqnarray}} \nc{\eeq}    {\end{eqnarray}}
\nc{\nn}      {\\\nonumber} \nc{\vs}      {\vspace{-0.275cm}}
\nc{\fra}    {\frac{1}{2}}
\nc{\mb}        {\mathbf}

\begin{document}

\preprint{}

\title{The hadronic equation of state of HESS J1731-347 from the relativistic mean-field model with tensor coupling}
\author{Kaixuan Huang}
\affiliation{School of Physics, Nankai University, Tianjin 300071,  China}

\author{Jinniu Hu}~\email{hujinniu@nankai.edu.cn}
\affiliation{School of Physics, Nankai University, Tianjin 300071,  China}
\affiliation{Shenzhen Research Institute of Nankai University, Shenzhen 518083, China}

\author{Ying Zhang}
\affiliation{Department of Physics, School of Science, Tianjin University, Tianjin 300354, China}

\author{Hong Shen}~\email{songtc@nankai.edu.cn}
\affiliation{School of Physics, Nankai University, Tianjin 300071,  China}

\date{\today}
\begin{abstract}
 A recent report has identified a central compact object (CCO) within the supernova remnant HESS J1731-347, with a mass and radius of $M=0.77^{+0.20}_{-0.17}M{\odot}$ and $R=10.4^{+0.86}_{-0.78}$ km, respectively. To investigate this light compact star, a density-dependent relativistic mean-field (DDRMF) model, specifically the DDVT model, has been employed. The DDVT model incorporates tensor couplings of vector mesons, which {can} successfully describe the properties of finite nuclei, such as charge radius, binding energy, and spin-orbit splitting. The introduction of tensor coupling reduces the influence of scalar mesons and generates a softer equation of state (EOS) in the outer core of the neutron star. Moreover, it has been found that the crust segment plays a crucial role in reproducing the mass-radius relation of HESS J1731-347, indicating a preference for a soft crust EOS. By manipulating the coupling strength of the isovector meson in the DDVT parameter set, a reasonable hadronic EOS has been obtained, satisfying the constraints from the gravitational-wave signal GW170817, the simultaneous mass-radius measurements from the NICER collaboration, and the properties of finite nuclei. Notably, the mass-radius relations derived from this hadronic EOS also accurately describe the observables of HESS J1731-347. Therefore, based on our estimation, the CCO in HESS J1731-347 may represent the lightest known neutron star.
\end{abstract}



\maketitle
\section{Introduction}
A neutron star is a compact object composed of extremely dense nuclear matter rich in neutrons. Understanding the properties of neutron stars involves solving the Tolman–Oppenheimer–Volkoff (TOV) equation~\cite{tolman1939,oppenheimer1939} with the equation of state (EOS) of neutron star matter as an input. Thus, the observation of neutron stars provides an excellent opportunity to study EOS under extreme conditions of density and isospin asymmetry. Over the years, the advancement of ground-based radio telescopes and the rapid development of space-based X-ray and $\gamma$-ray detection technology have facilitated numerous observations and studies of pulsars since the discovery of the first pulsar in 1967. These observations, in addition to the electromagnetic waves emitted by pulsars, have been augmented by the detection of gravitational wave signals by LIGO and Virgo collaborations, ushering in the era of multi-messenger astronomy.

Recent breakthroughs in neutron star observations have brought significant advancements. Firstly, the gravitational wave event GW170817, resulting from the merger of binary neutron stars, provided crucial information about binary masses and tidal deformability~\cite{abbott2017a, abbott2017b}. Secondly, through the use of pulse profile modeling of X-ray emissions from hot spots on the surface of isolated neutron star PSR J0030+0451~\cite{riley2019,miller2019} and the neutron star in the binary system PSR J0740+6620~\cite{riley2021,miller2021}, simultaneous measurements of masses and radii were achieved using the NICER (Neutron Star Interior Composition Explorer). These macroscopic measurements have significantly contributed to the constraints on the EOS of high-density neutron star matter.

In addition to observations and theoretical research focusing on massive neutron stars, there has been a growing focus on low-mass neutron stars, as they can provide valuable constraints on the EOS of dense matter at low-density region. Among the current mass measurements, the companion in PSR J0453+1559 is considered to be one of the lowest-mass neutron stars that has been detected, with a mass of $1.17\pm 0.004 M_{\odot}$ \cite{martinez2015}. High-mass X-ray binaries (HMXBs) are X-ray sources thought to host a compact object, typically a neutron star, which accretes material lost from its massive companion. Two eclipsing HMXB systems, 4U 1538-52 with a mass of $1.00\pm 0.10 M_{\odot}$ and SMC X-1 with a mass of $1.04\pm 0.09 M_{\odot}$ \cite{rawl2011}, are considered to be candidates for low-mass neutron stars. {Subsequent works have} aimed at achieving more precise determinations of neutron star masses in these two systems \cite{ozel2012, falanga2015}.

A low-mass neutron star can be formed directly from a supernova explosion through gravitational collapse triggered by electron capture. The mass range for such neutron stars is estimated to be around $1.15-1.2M_{\odot}$ \cite{nomoto1987,lattimer2012}. During the earliest stages of its evolution, a newborn hot proton-neutron star resulting from gravitational collapse is expected to have a minimum gravitational mass of approximately $0.89-1.13M_{\odot}$ \cite{strobel1999}. The mass of the neutron star can increase due to fallback after the explosion and accretion from binary companions \cite{lattimer2012}. Therefore, it is possible that a low-mass neutron star below the mentioned limits may not originate directly from a supernova explosion caused by gravitational collapse. Instead, it may involve other mechanisms that require further exploration.

Recently, a central compact object (CCO) was reported within the supernova remnant HESS J1731-347. The estimated mass and radius of this object are $M=0.77^{+0.20}_{-0.17}M_{\odot}$ and $R=10.4^{+0.86}_{-0.78}$ km, respectively \cite{doroshenko2022}. There are various speculations regarding its internal constitution. Some studies have focused on investigating it as a quark star \cite{clemente2022, horvath2023}, a hybrid star \cite{brodie2023}, or a neutron star \cite{doroshenko2022}.

The symmetry energy ($E_{\rm sym}$) and its density dependence play a crucial role in the EOS of neutron star matter, because of its highly isospin-asymmetric nature. These properties have significant implications for both astronomical observations and nuclear physics. The behavior of the symmetry energy near the nuclear saturation density ($\rho_0$) affects the structure of the neutron star crust \cite{oyamatsu2007,bao2014(89)} as well as the radius of neutron stars in the intermediate mass range \cite{lattimer2001}. In addition to their impact on neutron stars, $E_{\rm sym}$ and its slope ($L$) can be constrained by terrestrial experiments. Recent measurements of the neutron skin thickness ($R_{\rm skin}$) of $^{208}$Pb by PREX-I and PREX-II resulted in $R_{\rm skin}^{208}=0.283\pm 0.071$ fm \cite{abrahamyan2012, horowitz2012, adhikari2021}, leading to derived values of $E_{\rm sym}=38.1\pm4.7$ MeV and $L=106\pm37$ MeV based on the linear relation between $L$ and $R_{\rm skin}^{208}$ \cite{reed2021,roca-Maza2011,piekarewicz2021}. Similarly, the CREX collaboration reported the neutron skin thickness of $^{48}$Ca as $R_{\rm skin}^{48}=0.121\pm 0.026$ fm \cite{adhikari2022}, using the same method as PREX-II. Notably, the value of $L$ obtained from $^{48}$Ca is much smaller compared to that from PREX-II \cite{reinhard2022,zhang2022,lattimer2023}. A Bayesian analysis by Zhang and Chen inferred $E_{\rm sym}=30.2^{+4.1}_{-3.0}$ MeV and $L=15.3^{+46.8}_{-41.5}$ MeV at a $90\%$ confidence level \cite{zhang2022}, while Lattimer's work suggested that nuclear interactions optimally satisfying both measurements imply $L = 53 \pm 13$ MeV \cite{lattimer2023}. The substantial discrepancy between these two measurements poses significant challenges to the understanding of nuclear many-body methods.

The construction of an accurate EOS using efficient nuclear many-body methods plays a crucial role in bridging the gap between terrestrial experiments and astronomical observations, allowing for a comprehensive understanding of dense matter. The relativistic mean-field (RMF) model has been widely successful in describing the properties of finite nuclei and naturally extended to high-density regimes \cite{walecka1974,boguta1977,serot1979,sugahara1994,horowitz2001,meng2006}. In our study, we conducted a systematic investigation of the influence of symmetry energy on the neutron star crust \cite{bao2014(90)} and its mass-radius relation \cite{ji2019,hu2020} within the framework of the RMF model, employing two well-known parameterizations, namely TM1 \cite{sugahara1994} and IUFSU \cite{fattoyev2010}. The slope of symmetry energy ($L$) was controlled by adjusting the coupling constants of the isovector meson while fixing the symmetry energy $E_{\rm sym}$ at a density of 0.11 fm$^{-3}$. This choice of density is motivated by the fact that $E_{\rm sym}$ at approximately 0.11 fm$^{-3}$ is related to the binding energy of finite nuclei, rather than the saturation density \cite{zhang2013}. Our findings revealed that $L$ significantly affects the pasta structures and the properties of neutron stars in the intermediate mass range. Furthermore, we explored the density-dependent RMF (DDRMF) models \cite{typel1999}, which account for changes in the coupling strengths between mesons and nucleons within the nuclear medium, to describe massive neutron stars \cite{huang2020}. Notably, we discovered that a particular parameterization of the DDRMF model, known as DDVT, yields smaller radii for neutron stars in the low-mass region due to the inclusion of tensor coupling of vector mesons.

Our aim in this work is to construct the EOS for light neutron stars utilizing the density-dependent relativistic mean-field (DDRMF) model, incorporating the tensor coupling of vector mesons. This model not only reproduces the observables of finite nuclei in terrestrial experiments but also allows us to explore the possibility of HESS J1731-347 being a neutron star. This paper is structured as follows: Section \ref{sec2} presents the fundamental theoretical framework of the DDRMF model employed for the neutron star EOS. In Section \ref{sec3}, we delve into the impact of the symmetry energy within the DDRMF model on the properties of neutron stars, as well as the internal constituents of HESS J1731-347. Finally, in Section \ref{sec4}, we provide a conclusion summarizing our findings.

\section{The DDRMF model for the EOS of neutron star}\label{sec2}
Nucleons interact with each other through the exchange of scalar and vector mesons, such as $\sigma$, $\omega$, and $\rho$ mesons in the DDRMF model, where the coupling strengths between mesons and nucleons depend on the density, representing the nuclear medium effect. In this study, we choose the density-dependent parameter for the vector nucleon density, denoted as $\rho_N$. The density-dependent parameterizations for the $\sigma$ and $\omega$ mesons, denoted as $\Gamma_{\sigma N}$ and $\Gamma_{\omega N}$ respectively, are adopted in a fractional form, while the density-dependent parameterization for the $\rho$ meson, denoted as $\Gamma_{\rho  N}$, is in an exponential form. The Lagrangian of the DDRMF model, which includes the tensor coupling, can be expressed as,
   \begin{align}
	\mathcal{L}_{\rm DD}
	=&\overline{\psi}_ N\left[\gamma^{\mu}\left(i\partial_{\mu}-\Gamma_{\omega N}(\rho_ N)\omega_{\mu}-\frac{\Gamma_{\rho N}(\rho_N)}{2}\vec{\rho}_{\mu}\vec{\tau}\right)-\left(M_ N-\Gamma_{\sigma N}(\rho_N)\sigma\right)-\sigma^{\mu\nu}T_{\mu\nu}\right]\psi_N\nonumber\\
	&+\frac{1}{2}\left(\partial^{\mu}\sigma\partial_{\mu}\sigma-m_{\sigma}^2\sigma^2\right)
	-\frac{1}{4}W^{\mu\nu}W_{\mu\nu}+\frac{1}{2}m_{\omega}^2\omega_{\mu}\omega^{\mu}-\frac{1}{4}\vec{R}^{\mu\nu}\vec{R}_{\mu\nu}+\frac{1}{2}m_{\rho}^2\vec{\rho}_{\mu}\vec{\rho}^{\mu},
\end{align}
where $W_{\mu\nu}=\partial_\mu\omega_\nu-\partial_\nu\omega_\mu$ and $\vec{R}_{\mu\nu}=\partial_\mu\vec\rho_\nu-\partial_\nu\vec\rho_\mu$ are the asymmetric tensor fields. The $T_{\mu\nu}$ is the tensor coupling terms between the {vector mesons} and nucleons,
\begin{align}
	T_{\mu\nu}=\frac{\Gamma_{T\omega}(\rho_  N)}{2M_ N}W_{\mu\nu}+\frac{\Gamma_{T\rho}(\rho_  N)}{2M_ N}\vec R_{\mu\nu}\vec \tau.
	\end{align}
The inclusion of the tensor term in the DDRMF model has several effects. Firstly, it leads to a reduction in the mass of the $\sigma$ meson and introduces an additional attractive contribution in the finite nuclei system \cite{ddvt}. As a result, the coupling strength $\Gamma_{\sigma  N}$ is noticeably compressed in the DDRMF model with tensor coupling. Moreover, the tensor term also affects the spin-orbital splittings of finite nuclei. While the tensor coupling term does not contribute to infinite nuclear matter due to translation invariance, the reduction of the scalar meson field leads to a softer EOS and a larger effective nucleon mass at saturation density.

With the mean-field  and no-sea approximations, the energy density, $\varepsilon$ , and pressure, $P$, of infinite nuclear matter in DDRMF model can be written as \cite{huang2020},
\begin{align}
	\varepsilon=&\frac{1}{2}m_{\sigma}^2\sigma^2-\frac{1}{2}m_{\omega}^2\omega^2-\frac{1}{2}m_{\rho}^2\rho^2 
	+\Gamma_{\omega  N}(\rho_N)\omega\rho_N+\frac{\Gamma_{\rho N}(\rho_N)}{2}\rho\rho_3+\mathcal{E}_{\rm kin}^p+\mathcal{E}_{\rm kin}^n,\\\nonumber
	P=&\rho_N\Sigma_{R}(\rho_N)-\frac{1}{2}m_{\sigma}^2\sigma^2+\frac{1}{2}m_{\omega}^2\omega^2+\frac{1}{2}m_{\rho}^2\rho^2+P_{\rm kin}^p+P_{\rm kin}^n,
\end{align}
where the contributions from the kinetic energy are 
\begin{align}
	\mathcal{E}_{\rm kin}^i&=\frac{\gamma}{2\pi^2}\int_{0}^{k_{Fi}}k^2\sqrt{k^2+{M_i^{*}}^{2}}dk=\frac{\gamma}{16\pi^2}\left[k_{Fi}E_{Fi}^{*}\left(2k_{Fi}^2+{M_i^{*}}^2\right)+{M_i^{*}}^4{\rm ln}\frac{M_i^{*}}{k_{Fi}+E_{Fi}^{*}}\right], \\\nonumber
	P_{\rm kin}^i&=\frac{\gamma}{6\pi^2}\int_{0}^{k_{Fi}}\frac{k^4 dk}{\sqrt{k^2+{M_i^{*}}^{2}}}
	=\frac{\gamma}{48\pi^2}\left[k_{Fi}\left(2k_{Fi}^2-3{M_i^{*}}^2\right)E_{Fi}^{*}+3{M_i^{*}}^{4}{\rm ln}\frac{k_{Fi}+E_{Fi}^{*}}{M_i^{*}}\right].
\end{align}
The degeneracy of spin, is $\gamma=2$.

The nuclear matter in neutron stars, characterized by a large neutron fraction, exhibits significant isospin asymmetry. Therefore, the symmetry energy of nuclear matter and its density-dependent behavior play crucial roles. In the DDRMF model, the symmetry energy is defined as follows:
\begin{equation}\label{eq.esym}
	E_{\rm sym}=\frac{1}{2}\frac{\partial^2E/A(\rho_N,\delta)}{\partial \delta^2}\bigg|_{\delta=0}
	=\frac{k_F^2}{6E_F^*}+\frac{\Gamma_{\rho N}^2}{8m_{\rho}^2}\rho_N.
\end{equation}
where $E/A$ is the binding energy per nucleon and is determined by the energy density as $E/A=\varepsilon/\rho_N$ and $\delta=(\rho_n-\rho_p)/\rho_N$ is the asymmetry factor. The slope of the symmetry energy, $L$, which is related to the density-dependence of symmetry energy is derived as, 
\begin{equation}\label{eq.L}
	L=3\rho_{B}\frac{\partial E_{\rm sym}}{\partial\rho_N}=\frac{k_F^2}{3E_F^{*}}-\frac{k_F^4}{6E_F^{*3}}\left(1+\frac{2M_ N^{*}k_F}{\pi^2}\frac{\partial M_ N^{*}}{\partial\rho_N}\right)+
	\frac{3\Gamma_{\rho N}^2}{8m_{\rho}^2}\left(1-\frac{2a_{\rho}}{\rho_0}\rho_N\right)\rho_N.
\end{equation}

Upon observation, it is evident that the slope of symmetry is primarily determined by two factors: the coupling strength of the $\rho$ meson, denoted as $\Gamma_{\rho N}(\rho_N) = \Gamma_{\rho  N}(\rho_0) \exp(-a_\rho[(\rho_N/\rho_0)-1])$, and its density dependence. By adjusting the values of $\Gamma_{\rho N}(\rho_0)$ and $a_\rho$ within the DDVT parameterization, we have the ability to generate diverse values of $L$, while simultaneously maintaining the symmetry energy at a specific nucleon density along with other isoscalar properties of nuclear matter. Neutron star matter comprises both nucleons and leptons, which must adhere to charge neutrality and $\beta$-equilibrium conditions, thereby playing a significant role in determining the asymmetry factor.

\section{Numerical Results and Discussions}\label{sec3}
In our previous studies \cite{huang2020, huang2022}, we calculated the mass-radius relationships of neutron stars using different DDRMF parameter sets up until 2020. These parameter sets were obtained by fitting the ground-state properties of various finite nuclei. Our findings revealed that the DDVT set, which includes tensor coupling, resulted in a smaller radius for neutron stars in the lower mass region. Recently, several new DDRMF parameter sets based on the DD-MEX set, namely DD-MEX1, DD-MEX2, and DD-MEXY \cite{ddmex12Y}, have been proposed. The nuclear saturation properties, including the saturation density ($\rho_0$), binding energy per nucleon ($E/A$), symmetry energy ($E_{\rm sym}$), slope of symmetry energy ($L$), and effective nucleon mass ($M^*_N$), for the aforementioned parameterizations, are listed in Table \eqref{table.DDsets_sat}. Notably, DD-MEX2 exhibits a distinct slope of symmetry energy ($L$) compared to other DDRMF sets, while DDVT demonstrates a notable difference in effective nucleon mass ($M^*_N$). The fitting process of DD-MEX2 did not account for constraints from pure neutron matter and neutron skin, whereas the tensor coupling terms in DDVT suppress the magnitude of the $\sigma$ field, resulting in a larger effective nucleon mass. To examine the influence of effective mass on neutron star properties, we also considered three non-relativistic density-functional theory parameterizations, namely BSk19, BSk20, and BSk21 \cite{goriely2010, potekhin2013}, based on Skyrme-type effective interactions. These parameterizations have a relatively large effective mass of $0.8M_N$ at saturation density and slope of symmetry energy ($L$) values of $31.90$, $34.70$, and $46.60$ MeV, respectively.

\begin{table}[H] 
	\centering
	\caption{ Nuclear matter properties, such as the binding energy per nucleon $E/A$, incompressibility $K$, symmetry energy $E_{\rm sym}$, symmetry energy slope $L$, and effective mass $M_N^*/M_N$ at saturation density $\rho_0$ generated by different parameterizations.} \label{table.DDsets_sat}
	\begin{tabular}{r|cccccccccc}
		\hline \hline  
		SETS                    &$\rho_0 \rm[fm^{-3}]$  & $E/A$[MeV] & $E_{\rm sym}$[MeV] & $L$[MeV] & $M_N^*/M_N$   \\
		\hline   
		IUFSU \cite{fattoyev2010} & 0.1545 & -16.40 & 31.29 & 47.20  & 0.610   \\
		DDVT \cite{ddvt}          & 0.1536 & -16.92 & 31.56 & 42.35  & 0.667   \\   
		DDMEX \cite{ddmex}        & 0.1519 & -16.10 & 32.22 & 49.70  & 0.555   \\ 
		DDMEX1 \cite{ddmex12Y}    & 0.1510 & -16.01 & 31.78 & 53.23  & 0.570   \\
		DDMEX2 \cite{ddmex12Y}    & 0.1523 & -16.04 & 35.34 & 86.94  & 0.578   \\
		DDMEXY \cite{ddmex12Y}    & 0.1535 & -16.02 & 32.04 & 53.21  & 0.581   \\
		BSk19 \cite{goriely2010}  & 0.1596 & -16.08 & 30.00 & 31.90  & 0.800   \\  
		BSk20 \cite{goriely2010}  & 0.1596 & -16.08 & 30.00 & 34.70  & 0.800   \\ 
		BSk21 \cite{goriely2010}  & 0.1582 & -13.05 & 30.00 & 46.60  & 0.800   \\ 
		\hline\hline
	\end{tabular}
\end{table}

In the inner crust region of a neutron star, neutron-rich nuclei that are deformed exist within a mixture of neutron and electron gases, forming an inhomogeneous nuclear matter known as pasta structures. Our previous work \cite{ji2019} highlighted the crucial role of the EOS of the inner crust in determining the radius of neutron stars, particularly in the lower mass region. There is no inner crust EOS available for the DDVT parameter set. Consequently, we have adopted the inner crust EOS from the IUFSU set, which utilizes a self-consistent Thomas-Fermi (TF) approximation \cite{bao2014(90)}. This choice is justified by the similarity of their nuclear saturation properties. Similarly, for other DDRMF parameter sets, the inner crust EOS from the IUFSU set is also selected, while the inner crust EOSs have been calculated for the BSk19, BSk20, and BSk21 sets. Consequently, the corresponding unified EOSs are utilized for these three parameter sets.

\begin{figure}[htbp]
	\centering
	\includegraphics[scale=0.7]{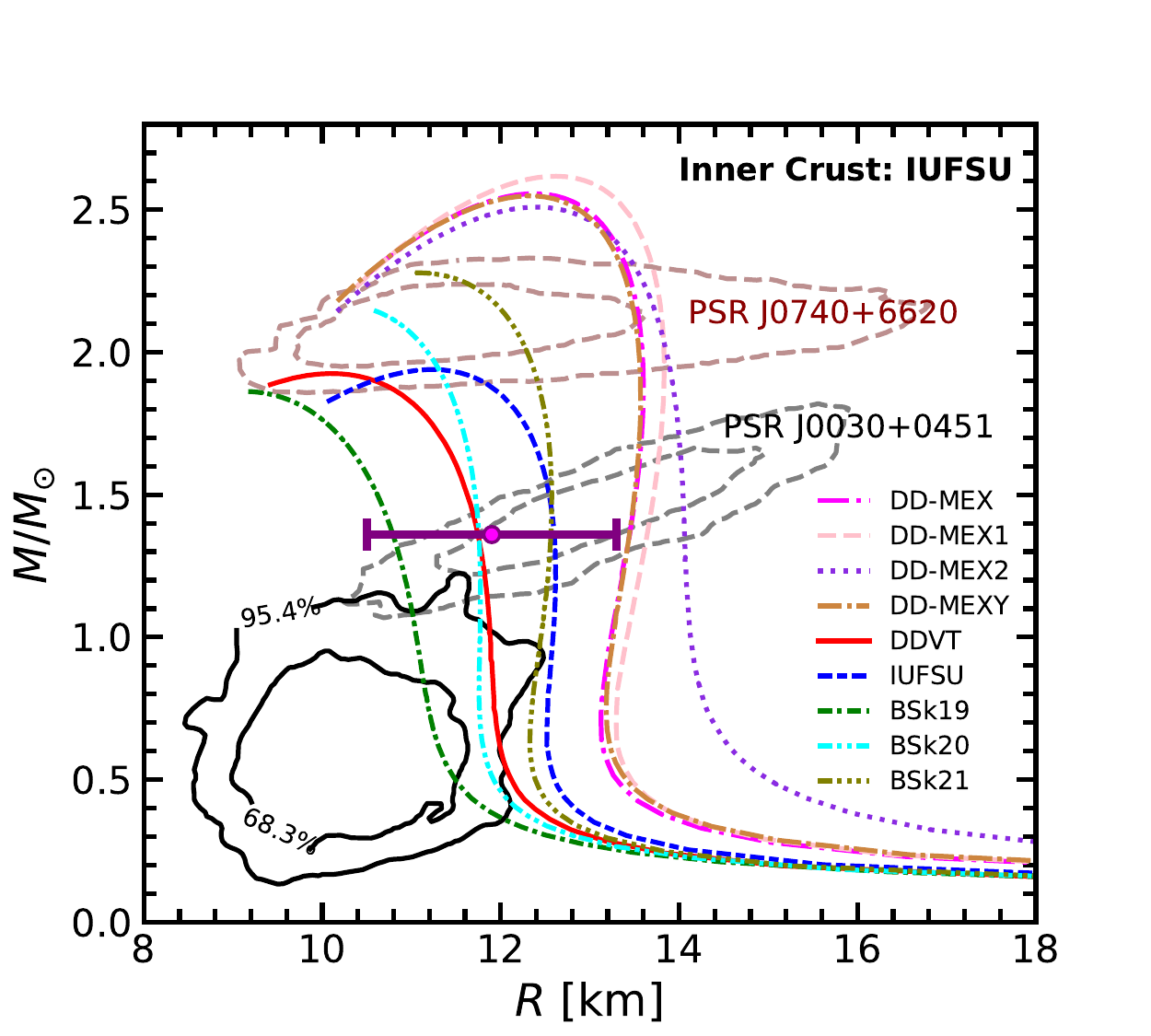}
	\caption{Mass-radius relations of neutron stars obtained using the EOSs from different DDRMF sets. The dotted contours show the 68.3\% and 95.4\% credibility mass-radius constraints from PSR J0740+6620 \cite{miller2021} and PSR J0030+0451 \cite{miller2019}. The solid contours represent the central compact objects within HESS J1731-347 \cite{doroshenko2022}. The horizontal error bar at 1.4$M_{\odot}$ is from GW170817 \cite{abbott2018}. } \label{fig.DDsets_MR}
\end{figure}

In Fig.~\ref{fig.DDsets_MR}, we present the mass-radius relations of neutron stars generated by the EOSs obtained from the aforementioned parameterizations. Additionally, we include mass-radius observations from measurements of PSR J0740+6620 and PSR J0030+0451 by NICER, as well as the compact central object (CCO) of HESS J1731-347, and the gravitational wave event from the merger of binary neutron stars, GW170817. As depicted in the figure, nearly all density functional theory parameter sets yield mass-radius relations that satisfy the $95.4\%$ confidence level constraints of PSR J0740+6620 and PSR J0030+0451. The existence of massive neutron stars above $2.5 M_\odot$ is attributed to the DD-MEX, DD-MEX1, DD-MEX2, and DD-MEXY sets, which exhibit very strong repulsive contributions from the $\omega$ meson. Among these sets, DD-MEX2 produces the largest radius of $1.4 M_\odot$ due to its highest values of $L$.

However, these four DDRMF parameter sets fail to accurately describe the measurement data from HESS J1731-347 \cite{doroshenko2022}. On the other hand, the EOSs from the DDVT and BSk series are capable of generating smaller radii in the lower mass region, which all fall within the $95.4\%$ confidence region of HESS J1731-347 \cite{doroshenko2022}. Remarkably, the mass-radius relation from BSk19 with $L=31.90$ MeV even aligns with the $68.3\%$ uncertainties of HESS J1731-347 {since BSk19 can provide a very soft EOS  compared to BSk20 and BSk21}. The GW170817 event provides a valuable constraint on the tidal deformability, which corresponds to a radius of a $1.4M_{\odot}$ neutron star within the range of $70<\Lambda_{1.4}<580$ and $R_{1.4}=11.9\pm1.4$ km, respectively \cite{abbott2018}. This constraint is represented in the figure by a horizontal error bar, favoring the soft EOSs from the DDVT, BSk19, BSk20, and BSk21 sets.

The properties of neutron stars, including the maximum mass ($M_{\rm max}$), the corresponding radius ($R_{\rm max}$), the central density ($\rho_c$), the radius at $1.4M_{\odot}$ ($R_{\rm 1.4}$), the dimensionless tidal deformability at $1.4M_{\odot}$ ($\Lambda_{\rm 1.4}$), the radius at $0.77M_{\odot}$ ($R_{0.77}$), and the corresponding central density ($\rho_{\rm 0.77}$) obtained from the aforementioned calculations, are listed in Table \ref{table.DDsets_MR}. The maximum masses of neutron stars from the IUFSU, DDVT, and BSk series fall within the range of $1.86-2.27M_\odot$, with corresponding radii all below $11.22$ km. The maximum radius is just $9.16$ km for the BSk19 set. However, these neutron stars exhibit very high central densities exceeding $6\rho_0$. 

Additionally, these sets generate smaller radii for neutron stars at the canonical $1.4M_\odot$ mass, resulting in lower tidal deformability values that align with the measurements from the GW170817 event. Notably, the BSk19 set produces a radius of only $11.17$ km at $0.77M_\odot$, which is comparable to the observable radius of HESS J1731-347. The central densities for neutron stars with a mass of $0.77M_\odot$ are approximately $2-3\rho_0$ for these relatively softer EOSs. In contrast, the stiffer EOSs from the DDMEX series yield very massive neutron stars above $2.5 M_\odot$, with relatively larger radii ranging from $12.34-12.59$ km. These neutron stars also exhibit larger radii and tidal deformabilities at $1.4 M_\odot$. However, these values do not agree with the measurements from HESS J1731-347.
 
\begin{table}[H] 
	\centering
	\caption{ Neutron star properties generated by different parameterizations.} 
	\label{table.DDsets_MR}
	\begin{tabular}{r|ccccccc}
		\hline \hline  
		SETS  & $M_{\rm max}[M_{\odot}]$ & $R_{\rm max}$[km]  & $\rho_c\rm[fm^{-3}]$ & $R_{1.4}$[km]  & $\Lambda_{1.4}$  & $R_{0.77}$[km]  & $\rho_{0.77}\rm[fm^{-3}]$  \\
		\hline  
		IUFSU  & 1.94 & 11.22 & 1.03 & 12.56 & 503 & 12.53 & 0.29  \\ 
		DDVT   & 1.93 & 10.08 & 1.22 & 11.72 & 306 & 11.93 & 0.34  \\
		DDMEX  & 2.56 & 12.37 & 0.78 & 13.47 & 788 & 13.12 & 0.24  \\ 
		DDMEX1 & 2.62 & 12.59 & 0.75 & 13.66 & 869 & 13.30 & 0.24  \\ 
		DDMEX2 & 2.51 & 12.39 & 0.79 & 14.06 & 896 & 14.28 & 0.23  \\ 
		DDMEXY & 2.55 & 12.34 & 0.78 & 13.46 & 787 & 13.19 & 0.24  \\   
		BSk19  & 1.86&  9.16 & 1.43 & 10.74 & 163 & 11.17 & 0.43  \\  
		BSk20  & 2.17 & 10.18 & 1.12 & 11.75 & 320 & 11.75 & 0.35  \\ 
		BSk21  & 2.27 & 11.05 & 0.97 & 12.59 & 522 & 12.37 & 0.29  \\ 
		\hline\hline
	\end{tabular}
\end{table}

We first aim to investigate the influence of different crust EOSs on the description of the light neutron star mass region. We considered crust EOSs obtained from the DDME$\delta$ model \cite{grill2014}, TM1 model, and IUFSU model \cite{bao2014(90)}. Our findings revealed the significant role played by the crust EOS in accurately characterizing the light neutron star mass range. To delve deeper into the effect of the crust, we constructed several neutron star EOSs by combining different crust EOSs while maintaining a consistent core EOS. The crust EOSs were derived from the IUFSU model at various values of $L$, with the symmetry energy fixed at $\rho_N=0.11~\rm fm^{-3}$ using the self-consistent Thomas-Fermi approximation \cite{bao2014(90)}. The core EOS was obtained from the DDVT set \cite{ddvt}. While the matching procedure of the crust-core transition has been shown to impact the properties of neutron stars \cite{doroshenko2022}, we simplified the approach by treating the intersection of these two EOSs, namely $(P-\varepsilon)$, as the crust-core transition point. The inner crust was described by the IUFSU models with $L=47,~60,~80,~110$ MeV, which are governed by the coupling strengths of the isovector meson. It was observed that a larger value of $L$ resulted in a lower crust-core transition density and led to a softer EOS for the inner crust, which is consistent with the findings of Ref. \cite{ji2019}. This can be easily understood through the expansion formula of the symmetry energy, $E_{\rm sym}(\rho_N)=E_{\rm sym}(\rho_0)+L(\rho_N-\rho_0)/3\rho_0+\cdots$.

\begin{figure}[htbp]
	\centering
	\includegraphics[scale=0.7]{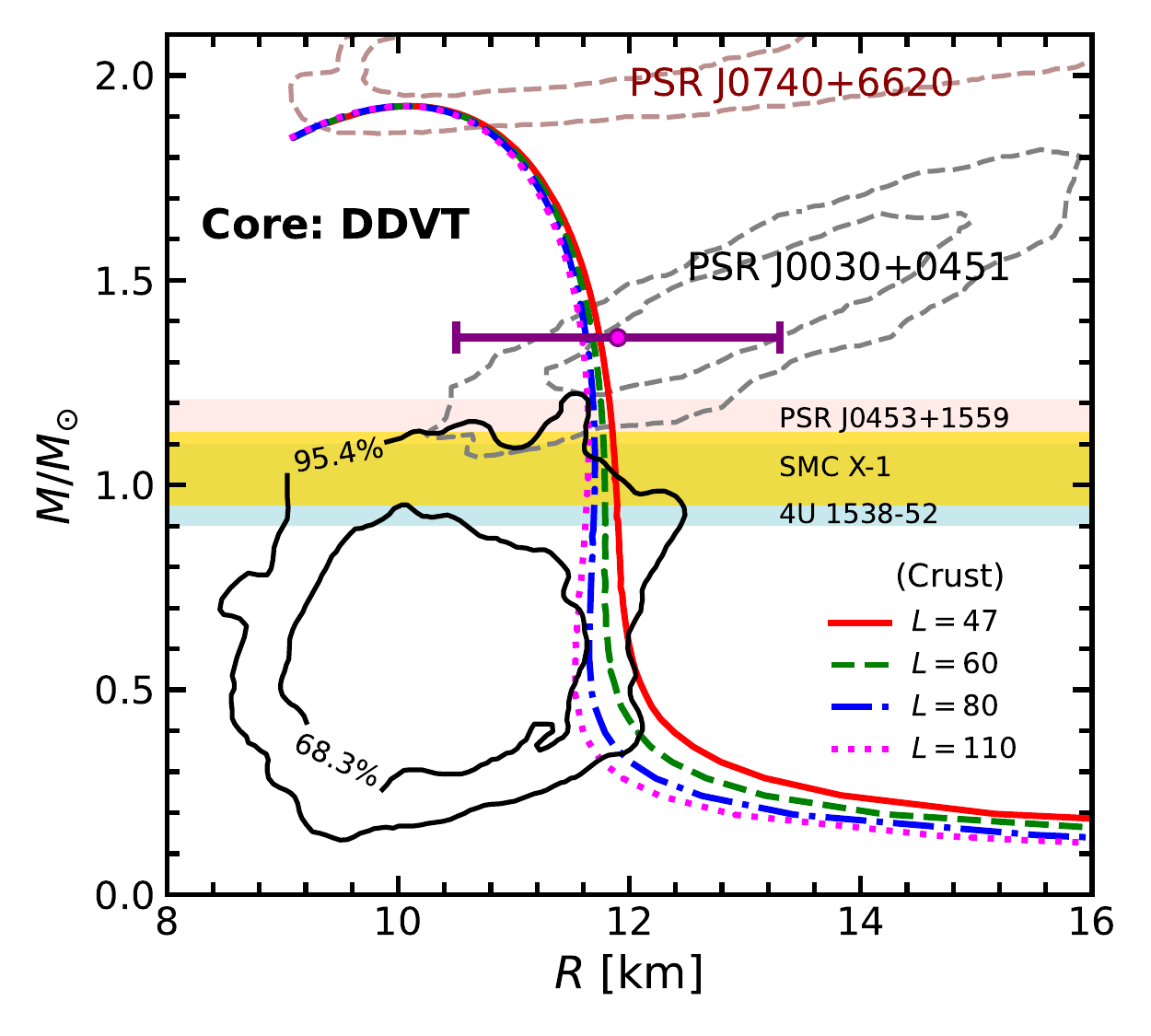}
	\caption{Mass-radius relations of neutron stars obtained using the constructed EOSs withing the crust EOSs from IUFSU models with $L=47,~60,~80,~110$ MeV and the core EOS from DDVT model. The dotted contours show the 68.3\% and 95.4\% credibility mass-radius constraints from PSR J0740+6620 \cite{miller2021} and PSR J0030+0451 \cite{miller2019}. The solid contours represent the central compact objects within HESS J1731-347 \cite{doroshenko2022}. The horizontal error bar at 1.4$M_{\odot}$ is from GW170817 \cite{abbott2018}.} \label{fig.DDVT(differ_crust)MR}
\end{figure}

In Figure \ref{fig.DDVT(differ_crust)MR}, we present the mass-radius relations obtained using the constructed EOSs in conjunction with crust EOSs from IUFSU models with $L=47$, $60$, $80$, and $110$ MeV, while the core EOS is based on the DDVT model. The crust EOS has minimal impact on the maximum mass of the neutron star and only slightly influences the radii at $1.4M_{\odot}$, which range from approximately $11.72$ km for $L=47$ MeV to $11.57$ km for $L=110$ MeV, resulting in a reduction of around $0.15$ km. The mass and radius of a CCO within the supernova remnant HESS J1731-347 have been estimated as $M=0.77^{+0.20}_{-0.17}M{\odot}$ and $R=10.4^{+0.86}_{-0.78}$ km, respectively \cite{doroshenko2022}, making the radius at $0.77M{\odot}$ an important factor to consider. As $L$ increases from $47$ MeV to $110$ MeV, the radius at $0.77M_{\odot}$ decreases from $11.93$ km to $11.59$ km, resulting in a reduction of approximately $0.35$ km. These diverse EOSs satisfy the $95.4\%$ confidence interval constraint of HESS J1731-347 since a higher $L$ value for the crust EOS yields a softer EOS, leading to smaller radii in the low mass region. Furthermore, the mass-radius relationship with $L=80$ and $110$ MeV crust EOSs at low densities even meets the $68.3\%$ credibility constraint established by HESS J1731-347.

Similarly, we also investigate the impact of the core EOS on neutron star properties. In this case, we continue to utilize the DDVT model as the core EOS due to its ability to yield relatively smaller radii in the low mass region, incorporating the tensor coupling term and satisfying the $95.4\%$ credibility constraint set by HESS J1731-347. Additionally, the DDVT model accurately describes the properties of finite nuclei. To achieve smaller radii in the intermediate mass region of neutron stars, such as $1.4M_{\odot}$ or lower mass regions, a smaller value of $L$ for the core EOS is required, which contrasts with the behavior of the crust EOS \cite{hu2020}. 

By adjusting $\Gamma_{\rho N}(\rho_0)$ and $a_{\rho}$ using Eq. \eqref{eq.esym} and Eq. \eqref{eq.L}, we can obtain core EOS configurations with reduced $L$ values, while maintaining $E_{\rm sym}=26.87$ MeV at $\rho_ B=0.11~\rm fm^{-3}$ and keeping the other isoscalar saturation properties unchanged, causing minimal impact on the binding energy of finite nuclei \cite{bao2014(90)}. In comparison to the original DDVT model, which yields $L=42.35$ MeV, we present several new DDVT sets for $L=26$, $30$, and $40$ MeV in Table \ref{table.differL}. It should be noted that when the slope of the symmetry energy falls below $L=26$ MeV, the speed of sound in nuclear matter becomes less than zero. The remaining coupling constants remain the same as those in the original DDVT model \cite{ddvt}. Table \ref{table.differL} also displays the $\rho$ meson coupling constants of the original DDVT model with $L=42.35$ MeV for comparison, revealing a clear decrease in the $\rho$ meson coupling strengths at nuclear saturation density, $\Gamma_{\rho N}(\rho_0)$, as $L$ is reduced.

\begin{table}[H] 
	\centering
	\caption{ Parameter $\Gamma_{\rho  N}(\rho_0)$ and $a_{\rho}$ generated from the DDVT model for different $L$ at the saturation point from $E_{\rm sym}$ fixed at $\rho_ B=0.11~\rm fm^{-3}$.} 
	\label{table.differL}
	\begin{tabular}{r|cccc}
		\hline \hline  
		$L$ [MeV]  &~~~ $\Gamma_{\rho N}(\rho_0)$ ~~~~~&$a_{\rho}$ \\
		\hline  
		26 & 7.250170  & 0.759445  \\
		30 & 7.367444  & 0.702782  \\
		40 & 7.637540  & 0.575940  \\
		42.35 & 7.697112  & 0.548702  \\ 
		\hline \hline
	\end{tabular}
\end{table} 

\begin{figure}[htbp]
	\centering
	\includegraphics[scale=0.7]{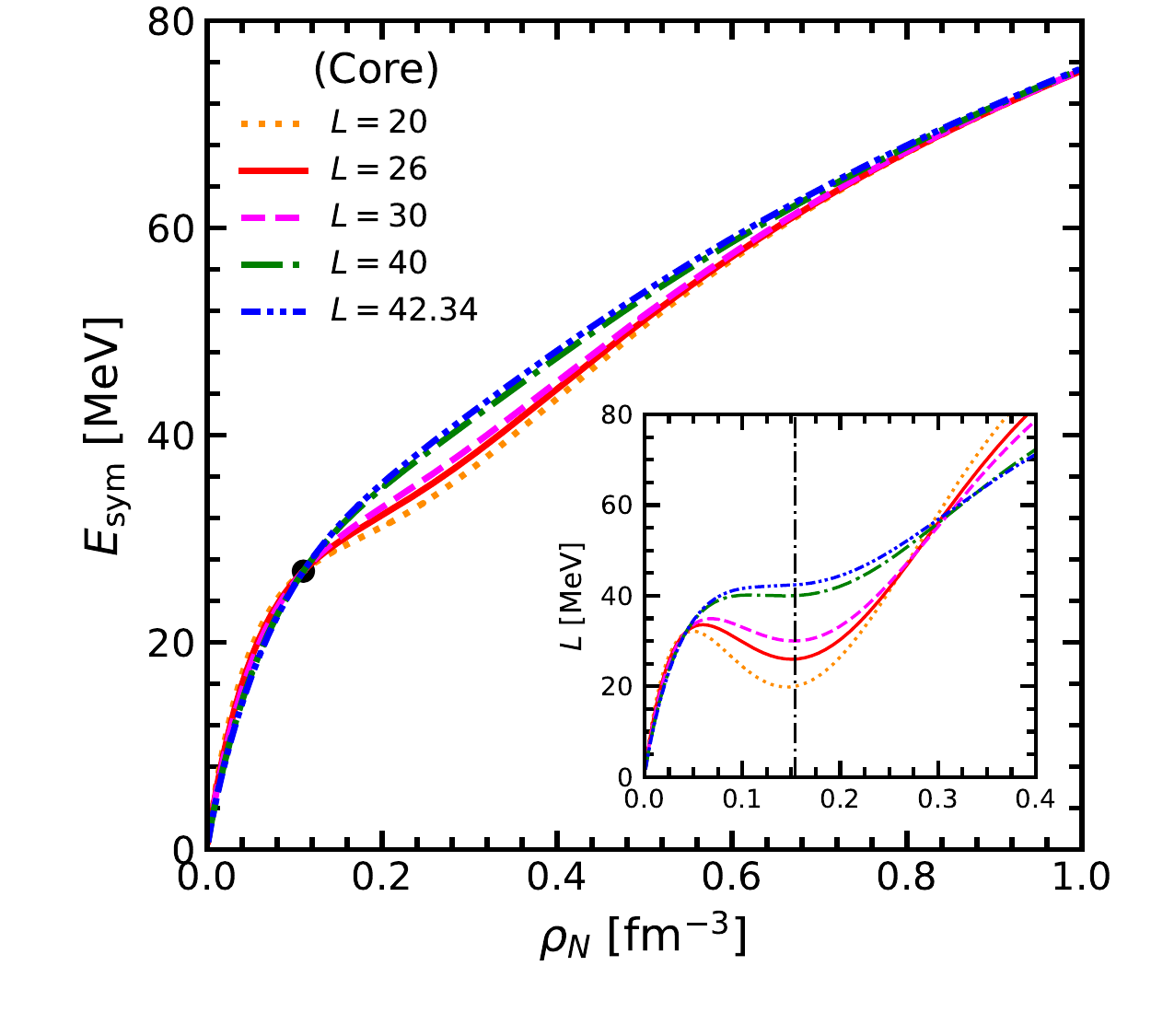}
	\caption{Symmetry energy $E_{\rm sym}$ as a function of the nucleon density for the DDVT sets with different $L$. The behaviors of the slope of symmetry energy $L$ are shown in the insert.} \label{fig.DDVT(differ_core)esymL}
\end{figure}

The density dependence of the symmetry energy $E_{\rm sym}$ is plotted in Figure \ref{fig.DDVT(differ_core)esymL} for the various $L$ parameter sets presented in Table \ref{table.differL}, which plays a crucial role in determining the properties of neutron stars. Smaller values of the $L$ parameter yield larger symmetry energy values below the sub-saturation density, while exhibiting smaller values in the high-density region. Unlike the behavior of $E_{\rm sym}$ obtained from nonlinear RMF parameter sets like the TM1 model in Ref. \cite{ji2019,hu2020}, the density-dependent model's $E_{\rm sym}$ converges above a density of 0.8 $\rm fm^{-3}$. This convergence is due to the influence of the {last term} in Equation \eqref{eq.esym}, which are related to the $\rho$ meson coupling constant, $\Gamma_{\rho N}(\rho_N)$. At high values of $\rho_N/\rho_0$, the magnitude of $\Gamma_{\rho N}(\rho_N)$ tends to approach zero exponentially. Consequently, the contribution of the {last term} gradually becomes negligible, and the symmetry energy at high densities is primarily determined by the contribution of the {first term}. The insert in the figure also displays the slopes of the symmetry energy as functions of density. Notably, these slopes undergo significant changes in the vicinity of nuclear saturation density when $L$ is very small.

\begin{figure}[htbp]
	\centering
	\includegraphics[scale=0.7]{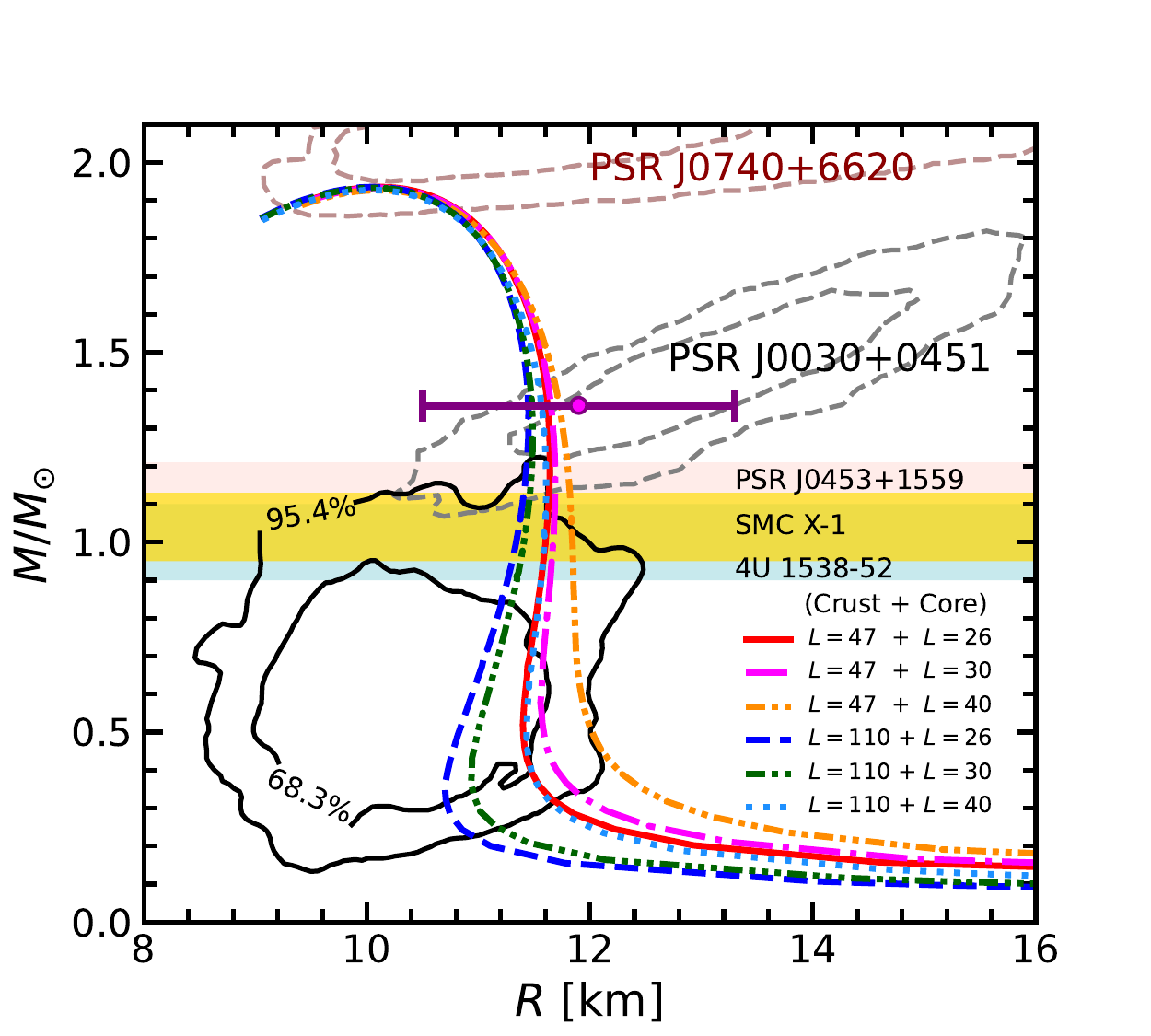}
	\caption{Mass-radius relations of neutron stars obtained using the sets from Table.\eqref{table.differL}. The dotted contours show the 68.3\% and 95.4\% credibility mass-radius constraints from PSR J0740+6620 \cite{miller2021} and PSR J0030+0451 \cite{miller2019}. The solid contours represent the central compact objects within HESS J1731-347 \cite{doroshenko2022}. The horizontal error bar at 1.4$M_{\odot}$ is from GW170817 \cite{abbott2018}.} \label{fig.DDVT(differ_core)MR}
\end{figure}

We will now utilize two extreme crust EOSs generated by the IUFSU family of models with $L=47$ and $110$ MeV \cite{bao2014(90)}, while the core EOSs will be obtained from the current family of DDVT parameterizations with $L=26$, $30$, and $40$ MeV, as listed in Table \ref{table.differL}. Each crust EOS is combined with three core EOSs, and the point of intersection between the two segments is considered as the crust-core transition point. By solving the Tolman-Oppenheimer-Volkoff (TOV) equation with these EOSs, we obtain the corresponding mass-radius relations for neutron stars, as depicted in Figure \ref{fig.DDVT(differ_core)MR}. It is observed that the $L$ parameter has minimal impact on the maximum mass and the corresponding radius of the neutron star. 

When considering the same crust EOS, the radius at $1.4M_{\odot}$ increases by approximately $0.1$ km as the core EOS changes from $L=26$ MeV to $L=40$ MeV, while the radius at $0.77M_{\odot}$ increases by $0.37-0.38$ km. This indicates that the $L$ parameter of the core EOS plays an equally important role in determining the radius of low-mass neutron stars, similar to the crust EOS in Figure \ref{fig.DDVT(differ_crust)MR}. However, the dependences of the two segments on $L$ are opposite. As previously demonstrated in Figure \ref{fig.DDVT(differ_crust)MR}, the EOSs combining the crust EOS with $L=47-110$ MeV and the original DDVT model already satisfy the $95.4\%$ credibility constraint of HESS J1731-347. Here, as the $L$ parameter of the core EOS decreases to $26-40$ MeV from the original value of $42.35$ MeV, the radii corresponding to the low-mass neutron stars obtained from the EOSs become sufficiently small to meet the $68.3\%$ credibility constraint, except for the combination of $L=47$ MeV for the crust EOS and $L=40$ MeV for the core EOS. Therefore, the observational data from HESS J1731-347 suggest the requirement of a crust EOS with a higher $L$ parameter and a core EOS with a lower $L$ parameter, representing an extremely soft EOS in both segments which is also consistent with the observables of PSR J0740+6620, PSR J0030+0451 from NICER, the GW170817 event, and the PREXII. This explains why the BSk19 EOS can also accurately describe HESS J1731-347.

Finally, we present the properties of neutron stars for the aforementioned EOSs in Table \ref{table.differL_MR}. These properties include the maximum mass ($M_{\rm max}$), the corresponding radius ($R_{\rm max}$), the central density ($\rho_c$), the radius ($R_{\rm 1.4}$), and the dimensionless tidal deformability ($\Lambda_{\rm 1.4}$) at $1.4M_{\odot}$, as well as the radius ($R_{\rm 0.77}$) and the corresponding density at $0.77M_{\odot}$. The maximum masses, corresponding radii, and central densities are nearly identical for all six EOSs. However, as the mass decreases, the differences in radius become more significant. The radius at $0.77M_{\odot}$ exhibits a disparity of $0.63$ km between the softest and stiffest EOSs, which results in the former satisfying the observables of HESS J1731-347 at the $68.3\%$ confidence level. Furthermore, all the tidal deformabilities ($\Lambda_{\rm 1.4}$) at $1.4M_{\odot}$ comply with the analysis by LIGO and VIRGO, with $\Lambda_{\rm 1.4}=190_{-120}^{+390}$ obtained from GW170817 \cite{abbott2018}.
 
\begin{table}[H] 
	\centering
	\caption{Neutron star properties generated by the sets from Table. \ref{table.differL}.} \label{table.differL_MR}
	\begin{tabular}{c|c|ccccccccc} 
		\hline \hline  
		Crust &Core  & $M_{\rm max}[M_{\odot}]$ & $R_{\rm max}$[km]  & $\rho_c\rm[fm^{-3}]$ & $R_{1.4}$[km]   & $\Lambda_{1.4}$ & $R_{0.77}$[km]  & $\rho_{0.77}\rm[fm^{-3}]$  \\
		\hline
		\multirow{3}*{$L=47$}  
		&$L=26$ & 1.93 & 10.10 & 1.23 & 11.61  & 316 & 11.50 & 0.34  \\
		&$L=30$ & 1.93& 10.11 & 1.23 & 11.63  & 312 & 11.60 & 0.34  \\ 
		&$L=40$ & 1.93 & 10.12 & 1.23 & 11.70  & 307 & 11.87 & 0.34  \\ 
		\hline
		\multirow{3}*{$L=110$}  
		&$L=26$ & 1.93 & 10.04 & 1.23 & 11.44 & 321 & 11.14 & 0.34  \\ 
		&$L=30$ & 1.93& 10.04 & 1.23 & 11.47 & 318 & 11.24 & 0.34  \\  
		&$L=40$ & 1.93 & 10.06 & 1.23 & 11.55 & 308 & 11.52 & 0.34  \\ 
		\hline \hline
	\end{tabular}
\end{table}

\section{Conclusions} \label{sec4}
In this study, we investigated the hadronic EOS within the framework of the density-dependent relativistic mean-field (RMF) model to describe low-mass neutron stars. We incorporate tensor coupling terms between vector mesons and nucleons, referred as DDVT, which reduce the magnitude of the scalar field and result in a soft EOS and a large effective nucleon mass. The DDVT model also demonstrates excellent agreement with the properties of finite nuclei. By fixing the crust EOS obtained by the IUFSU parameterization set with $L=47$ MeV, and utilizing the core EOS from DDVT with {$L=42.35$} MeV, the mass-radius relation derived from this combined EOS satisfies the $95.4\%$ confidence constraint of HESS J1731-347.

Initially, we examined the influence of the crust EOS while maintaining a fixed core EOS from the DDVT set. When transitioning from a crust EOS with $L=47$ MeV to $L=110$ MeV, the radius of a $1.4M_{\odot}$ neutron star decreased from approximately $11.72$ km to $11.57$ km, resulting in a reduction of about 0.15 km. Similarly, at $0.77M_{\odot}$, the radius decreased by approximately $0.35$ km. This effect arises due to the negative role played by $L$ in the density below the nuclear saturation point, as defined by the slope of the symmetry energy. Therefore, a larger $L$ interaction leads to a softer crust EOS.

To achieve a softer EOS in the high-density region, a family of parameter sets based on the DDVT model was developed by manipulating the coupling constants of the $\rho$ meson. By selecting a smaller $L$ at nuclear saturation density and fixing the symmetry energy at $0.11$ fm$^{-3}$, while keeping the other coupling strengths unchanged, we successfully maintained the isoscalar nuclear saturation properties and the binding energy of finite nuclei. Consequently, we obtained the DDVT series sets with $L=26,~30$, and $40$ MeV, which were used to calculate the core EOSs of the neutron star.

By fitting the crust EOS and utilizing the core EOSs generated by the DDVT family set with $L=26$ MeV, the radius at $0.77M_{\odot}$ decreases by approximately $0.37$ km compared to the case with $L=40$ MeV. The largest discrepancy in radius at $0.77M_{\odot}$ between the softest and stiffest EOSs across the entire density range amounts to {$0.73$} km, and the former satisfies the $68.3\%$ credit constraint set by HESS J1731-347. Therefore, if the central compact object (CCO) of HESS J1731-347 is a neutron star composed of hadronic matter, its EOS should exhibit significant softness throughout the density range, achieved by adopting a core with smaller $L$ and a crust with larger $L$. The mass-radius relation from this EOS also satisfies recent observables from the NICER and gravitational wave detection.

\section{Acknowledgments}
This work was supported in part by the National Natural Science Foundation of China	 (Grant  Nos. 11775119 and 12175109), and the Natural Science Foundation of Tianjin (Grant  No: 19JCYBJC30800).

\end{document}